\documentclass[conference]{IEEEtran}

\usepackage{cite}
\usepackage{amsmath,amssymb,amsfonts}
\usepackage{algorithmic}
\usepackage{graphicx}
\usepackage{textcomp}
\usepackage{xcolor}
\def\BibTeX{{\rm B\kern-.05em{\sc i\kern-.025em b}\kern-.08em
T\kern-.1667em\lower.7ex\hbox{E}\kern-.125emX}}

\newcommand{\bff}{{\mathbf f}}
\newcommand{\bk}{{\mathbf k}}
\newcommand{\br}{{\mathbf r}}
\newcommand{\bv}{{\mathbf v}}

\newcommand{\dC}{{\mathbb C}}
\newcommand{\dR}{{\mathbb R}}

\newcommand{\vr}{\mbox{$\bf r $}}
\newcommand{\vc}{\mbox{$\bf c $}}
\newcommand{\vw}{\mbox{$\bf w $}}

\newcommand{\be}{\begin{equation}}
\newcommand{\ee}{\end{equation}}
\newcommand{\bea}{\begin{eqnarray}}
\newcommand{\eea}{\end{eqnarray}}

\newcommand{\MYfooter}{\smash{
\hfil\parbox[t][\height][t]{\textwidth}{\centering
\thepage}\hfil\hbox{}}}

\makeatletter

\def\ps@headings{%
\def\@oddhead{\parbox[t][\height][t]{\textwidth}{\centering
Accepted for presentation at IEEE ICC 2022 Workshop  on Wireless Propagation Channels for 5G and B5G, \textcopyright 2022 IEEE
}}%
}

\def\ps@IEEEtitlepagestyle{%
\def\@oddhead{\parbox[t][\height][t]{\textwidth}{\centering
Accepted for presentation at IEEE ICC 2022 Workshop  on Wireless Propagation Channels for 5G and B5G, \textcopyright 2022 IEEE
}\hfil\hbox{}}%
\def\@evenhead{\scriptsize\thepage \hfil \leftmark\mbox{}}%
\def\@evenfoot{\MYfooter}}

\makeatother
\begin{document}

\title{Deep Learning for Wireless Dynamics}

\author{
\IEEEauthorblockN{Heunchul Lee, Jaeseong Jeong, Zhao Wang,}
\IEEEauthorblockA{\textit{Ericsson Research,}
\textit{Ericsson AB}\\
Stockholm, Sweden \\
\{heunchul.lee,jaeseong.jeong,zhao.wang\}@ericsson.com}\\
}
\maketitle

\begin{abstract}
This paper aims to predict radio channel variations over time by deep learning from channel observations without knowledge of the underlying channel dynamics. 
In next-generation wideband cellular systems, multicarrier transmission for higher data rate leads to high-resolution predicting problem. 
By leveraging recent advances of deep learning in high-resolution image processing, we propose a purely data-driven  deep learning (DL) approach to predicting high-resolution temporal evolution of wideband radio channels. 
In order to investigate the effect of architectural design choices, we develop and study three deep learning prediction models, namely, baseline, image completion and next-frame prediction models using UNet.
Numerical results show that the proposed DL approach achieves  a 52\% lower prediction error than the traditional approach based on Kalman filter (KF) in mean absolute errors. 
To quantify impact of channel aging and prediction on precoding performance, we  also evaluate the performance degradation due to outdated and predicted channel state information (CSI) compared to perfect CSI.  
Our simulations show that  the proposed DL approach can reduce the performance loss due to channel aging by 71\% through adapting precoding vector to changes in radio channel while the traditional KF approach only shows a 27\% reduction.
\end{abstract}


\section{Introduction}
As we are moving towards the sixth generation (6G) mobile communications system, artificial intelligence (AI) has the potential to enhance the systems from different aspects, compared to traditional optimization approaches.
In this regard, a new work item on AI for air interface has been recently approved for Release 18 at the 3rd Generation Partnership Project (3GPP) radio access network (RAN) plenary meeting \cite{3gpp:ai}. 
Among the different features described therein for enhancing the system performance via AI, we are particularly interested in channel prediction arising in use cases of channel state information (CSI) feedback, beam management and positioning processes.
In this paper, we aim to study the problem of channel prediction over time. 

Radio channel represents the relation between the transmitted and received signals that captures not only physical wave propagation between the transmitter and the receiver but also antenna configuration at both link ends.
Ideally, an accurate channel prediction requires a model for the temporal channel evolution. 
Physical (propagation) channel can be viewed as a nonlinear dynamical system and can be represented by partial differential equations such as Maxwell's equations.
However, the true channel model may not be known due to the underlying complex propagation phenomenon that can not be captured by using the domain knowledge and, even if it is known, the model may not enable computationally efficient methods for predicting channel evolution in time. This paper focuses on data-driven channel prediction by deep learning from channel data, instead of using domain knowledge to explicitly build prediction models. 

In the wideband cellular systems such as the fifth generation (5G) new radio (NR), multicarrier transmission for higher data rate leads to high-resolution predicting problem. 
For the slot-based 5G system, a slot consists of a number of OFDM symbols in time domain. 
In frequency domain, one OFDM symbol is constructed of subcarriers. 
Therefore, the channel state $h_k$ at a slot $k$ can by represented by the values of channel responses on the time-frequency grid of size equal to $T\times F$, where $T$ is the number of OFDM symbols and $F$ is the number of total subcarriers. 
Time series of states $h_k$ from data measurement provide valuable information about the underlying channel dynamics of our wireless communication system. 
So, in the absence of the model, time-varying radio channel can be described by using an alternative description of its dynamical system in a state-evolution function with a memory  
\bea \label{eq:pre}
h_{k+1}=\bff(h_{k},h_{k-1},...,h_{k-m+1}),
\eea
where $m$ is the memory size.
The exact evolution function $\bff$ in \eqref{eq:pre} is generally unknown and nonlinear.
The main contribution of this paper is to demonstrate that the temporal evolution of radio channel can be successfully predicted through function approximation to the evolution function.

Recent works have demonstrated  the successful application of convolutional network architectures in many image processing tasks involving high-resolution data \cite{UNet:15} \cite{pix2pix:17}. 
In this work, we use the convolutional network architecture proposed in \cite{UNet:15} as a function approximation for the evolution function in \eqref{eq:pre}. 
In particular, in order to investigate the effect of architectural design choices, we develop and study three prediction models, namely, baseline, image completion and next-frame prediction models using autoencoder (AE) and UNet. 
We evaluate the performance of the three prediction models over synthetic channel data generated by using the 3GPP channel model as well as real channel data obtained from measurement campaigns in real-world environments. Both datasets consist of high-resolution images corresponding to the time series of simulated or measured channel states. 

First, we present and evaluate the three UNet-based prediction models on synthetic channel data in comparison with the AE-based models.
Numerical results show that the UNet can improve channel prediction performance up to 50 times in mean absolute errors, demonstrating superior accuracy of the UNet architecture over the AE.
Then, for validation on real channel data, the UNet-based next-frame prediction will be evaluated by using real channel data obtained from measurement campaign in real-world outdoor environment. 
We observe that the proposed approach is capable of predicting accurate channels in real data. 
In order to verify the accuracy, we also compare our deep learning (DL) approach with a traditional approach based on Kalman-Filter (KF). 
It is shown that the proposed DL approach achieves 52\% lower prediction error than the traditional KF approach.
To quantify impact of channel prediction accuracy on precoding performance, we have further evaluated the data-driven approach in terms of outage capacity.
Simulation results demonstrate that the proposed channel prediction approach can be applied to adapt the precoding vector to changes in radio channel. 

The remainder of this paper is organized as follows. 
Section \ref{sec:1} briefly presents channel model and provides an overview of channel data used for our work. 
In Section \ref{sec:2}, we describe channel prediction formulation and three channel prediction models.
In Section \ref{sec:3}, we provide simulation results and comparisons with the traditional KF approach.
Finally, conclusions are made in Section \ref{sec:4}.

\section{Channel model and data} \label{sec:1}
In this section, we briefly describe the radio channel model given by a multi-path propagation model in a wireless scenario, where each path has large- and small-scale features from pathloss and shadowing to channel dispersion and fluctuation behavior due to multipath and Doppler spread. We also provide an overview of our datasets that have been collected in simulation and real-world system.

\subsection{Channel model}
Channel model plays an important role in developing and evaluating wireless communication systems. 
{\it Radio channel} describes the relation between the transmitted and received signals. 
Radio channel model comprises two parts: {\it physical channel} that captures physical wave propagation between transmitter and receiver and {\it antenna array model} that captures antenna configuration at both link ends \cite{Asplund:20}. 
In other words, physical channel focuses on wave propagation between the location of transmitter and receiver without taking the antenna array into account while antenna array model captures the impact of transmitting and receiving antenna through antenna radio patterns. 
As wireless systems have evolved to multi-antenna wideband systems using considerably higher radio frequencies, 
modelling of physical channel and antenna array has attracted much attention and as a result, channel models have been developed and standardized within wireless standards such as the 3GPP \cite{3gpp:138.901}.

Physical channel depends on surrounding environments including buildings, structure of roads and trees or foliage.
The radio waves interact with these various objects through processes such as reflection, transmission, diffraction, and scattering. 
The physical channel model can be described by the propagation paths. Each path $n$ can be characterized by a set of properties including a time delay, denoted by $\tau_n$, a path loss by $a_n$, propagation directions of departure from the transmitter and arrival at the receiver by the unit vector $\bk_{t,n}$ and $\bk_{r,n}$, and spatial positions by $\vr_t$ and $\vr_r$ of the transmitter and the receiver, respectively.

In general, a time-varying wideband channel can be modeled approximately by a transfer function of baseband frequency $f$, time $t$, and spatial position as \cite{Asplund:20}
\bea \label{eq:h}
H(f,t,\vr_t,\vr_r) \triangleq \sum_{n}a_n a_{tr} e^{-2\pi f_c \tau_n} e^{-2\pi f \triangle\tau_n}\times\nonumber \\ 
e^{j \bk_{t,n}\cdot \br_t} e^{j \bk_{r,n}\cdot \br_r}e^{j \bk_{r,n}\cdot \bv_r t},
\eea 
where $a_{tr}$ captures the received signal strength due to the transmit and receive antenna configuration, $f_c$ is a carrier frequency, $e^{-2\pi f_c \tau_n}$ represents the phase shift due to the propagation path length, $e^{-2\pi f \triangle\tau_n}$ denotes the phase shift due to variations of the frequency $f$ and relative time delay difference $\triangle\tau_n$, and the remaining terms represent the phase shifts due to the relative spatial positions or variations of receiver position due to the vector velocity $\bv_r$ over time $t$. 

The frequency-domain channel representation of radio channel in \eqref{eq:h} can be converted to an equivalent channel model in time domain. 
Furthermore, the time-domain channel model can be extended by incorporating dual-polarization and multi-antenna modeling. 
The resulting dual-polarized multi-input multi-output (MIMO) spatial channel model (SCM) is widely standardized and used for simulations of wireless communication.

\subsection{Channel data}
In this subsection, we present two datasets used for performance evaluation: one from synthetic channel and the other from real measurements.
The same antenna configuration is considered for both synthetic and real channel data.
Base station (BS) antenna array is deployed by a $\pm 45^{\circ}$ dual-polarized 4-by-1 (vertical-by-horizontal) antenna array with vertical antenna spacing $0.7 \lambda$, where $\lambda$ is the carrier wavelength.
The electrical downtilt is set as $6^{\circ}$ for the one-column array. 
User equipment (UE) uses a single antenna polarized within the horizontal plane. This 8-port BS and 1-port UE antenna setup results in 8x1 multi-input single-output (MISO) downlink systems. 

We generate synthetic data by using the three dimensional (3D) SCM model developed by 3GPP \cite{3gpp:138.901} and associated parameters. We assume 600 subcarriers and
carrier frequency 3.5GHz with subcarrier spacing 15kHz. In this assumption, each slot has duration of 1ms.
Channel coefficients are generated as follows. Urban macro (UMa) model is specified as simulation environment. 
3D locations of BS and UEs are determined through a sequence of drops, where a drop is defined as one simulation run using the same user parameters according to the specified simulation environment, including large scale parameters and small scale parameters.
The large scale parameters include delay spread, angular spreads, Rician K factor and shadow fading. The small scale parameters such as delays and arrival angles and departure angles are drawn randomly from the distribution defined in \cite{3gpp:138.901}. For UE mobility model, all UEs are deployed as outdoor user at speed 15km/h. Time-domain channel coefficients are generated for each transmitter and receiver element pair according to Equation (7.5-30) in \cite{3gpp:138.901}. 
Pathloss and shadowing are applied for the channel coefficients.
It's worth mentioning that our prediction models learn to predict channel at a fast fading scale while slow fading caused by path loss due to distance and shadowing effects remains constant within the time scale of our interest.
The time-domain channel coefficients are converted to channel responses in frequency domain and used to determine the channel state $h^n_k \in \dR^{T \times F}$ on time-frequency radio resource grid of OFDM symbols on one axis and subcarriers on the other axis.
Finally, we obtain synthetic data samples $s_i$ that includes channel responses from $(m+1)$ consecutive slots, i.e., $s_n=\left[h^n_{k-m+1} ,\cdots, h^n_{k},h^n_{k+1}\right]$ where each sample $s_n$ is normalized by the maximum absolute value of all the elements on a sample basis so that the value range is bound by -1 and 1. 
The first $(m-1)$ states, $h^n_{k-m+1} ,\cdots, h^n_{k}$, in each sample $s_n$ are used as a conditioning input for the prediction models to estimate the last state $h^n_{k+1}$.  
In our synthetic channel data, $F$ is chosen to be equal to the number of total subcarriers and
$T$ is set to be the same as  the number of OFDM symbols per slot  in the normal cyclic prefix mode, i.e.,  $F$=600 and $T$=14. 
Under the assumptions presented above in this section, the coherence time at the limit of auto-correlation function above 0.5 is approximately estimated to $3.68$ms, which is about four slots of length 1ms. In the rest of this paper we use the memory size $m=4$.

Real channel data are obtained from measurement campaign in an urban macro-cell scenario in Sweden. 
The macro BS is deployed with the antenna model as specified in the beginning of this section, while a testbed UE with a horizontally polarized omni-directional antenna is implemented on a moving vehicle. 
The testbed UE vehicle moves at an average speed of 9km/h during the whole measurement campaign. 
The deployed BS and UE are connected via a typical 5G NR system with subcarrier spacing 30kHz in the 4.9 GHz time-division duplexing (TDD) band.
The outdoor propagation measurements of UE-to-BS channel were conducted at the BS based on sounding reference signals (SRS) transmission in uplink.
A periodic SRS is transmitted from the UE to the BS with a periodicity 5ms, corresponding to the duration of ten slots.
SRS is mapped to every fourth subcarrier, which corresponds to a frequency interval of 120kHz. 
According to the geographical information of the driving route and detailed measurement analysis, we expect that approximately 70\% of the route are non-line-of-sight (NLOS), which offers a great opportunity for validating the performance of channel prediction models. 
Meanwhile, the target signal-to-noise ratio (SNR) on the SRS resource elements is set to 30dB in order to maintain the measurement quality. 
In our real channel data, 
the frequency resolution $F$ of channel image at measured state $k$ is the same as that in the synthetic data, but  the time resolution $T$ is reduced to be one due to the limitation in the standard.

\section{Prediction formulation and models} \label{sec:2}
In this section, we describe channel prediction formulation and three channel prediction models including  baseline, image completion and next-frame prediction models.

\subsection{Prediction formulation}
The goal is to predict future channel state $h_{k+1}$ given previous channel states, $h_{k-m+1}$ ,$\cdots$, $h_{k}$. 
Figure \ref{f3} depicts an illustrative example of input and output state pair with $m=4$ in synthetic channel data, where input and output state pairs can be represented by high-resolution images of size $T\times F$.
The conditioning channel images, composed from the past four slots $k-3$ to $k$, are depicted by a black box and the target channel image, given by next slot $k+1$, is depicted by a red box.
The temporal evolution of channel state can be estimated by deep learning from the underlying patterns over the sequence of channel images. 
In this work, we consider a deep learning model $\bff_{\theta}$ with parameters ${\theta}$ as a function approximation for the evolution function in \eqref{eq:pre} as follows:
\bea \label{eq:pre2}
h_{k+1}=\bff_{\theta}(h_{k},h_{k-1},...,h_{k-m+1}).
\eea

The pixel-wise metric $l_1$ is a widely accepted metric for defining loss function in image processing tasks.
As a measure of prediction accuracy, we will use the same metric for quantifying the prediction performance. 
The corresponding loss function is defined by mean absolute error (MAE) on all the pixels between the predicted channel image and the ground truth as follows
\bea \label{eq:l1}
l_1 = \frac{1}{N\times T\times F}\sum_{n=1}^{N}\sum_{t=1}^{T}\sum_{f=1}^{F} \left| h_{k+1}^{n}(t,f)-\hat{h}_{k+1}^{n}(t,f)\right|,
\eea
where $h_{k+1}^{n}(t,f)$ and $\hat{h}_{k+1}^{n}(t,f)$ denote the $(t,f)$th element of ground truth and predictions, respectively, and $N$ is the size of mini-batch (or batch). 
We note that $h_{k+1}^{n}(t,f)$ corresponds to the real- or imaginary-part of channel response at the $f$th subcarrier in frequency domain on the $t$th OFDM symbol in time domain within a given slot $k+1$. This means that our proposed prediction models will predict the real- and imaginary-parts of channel responses independently of each other.

\subsection{Prediction models}
In this subsection, we present three predictions models: baseline model denoted by $\bff_{\theta}^{bs}$, image completion model by $\bff_{\theta}^{ic}$ and next-frame prediction model by $\bff_{\theta}^{nf}$ that learns to predict the next channel state $h_{k+1}$ given the channel states $h_{k},h_{k-1},...,h_{k-3}$, i.e. $m=4$.

\begin{figure}
\centerline{\includegraphics[trim=1cm 1.0cm 1.5cm 1.3cm,clip,width=0.7 \linewidth]{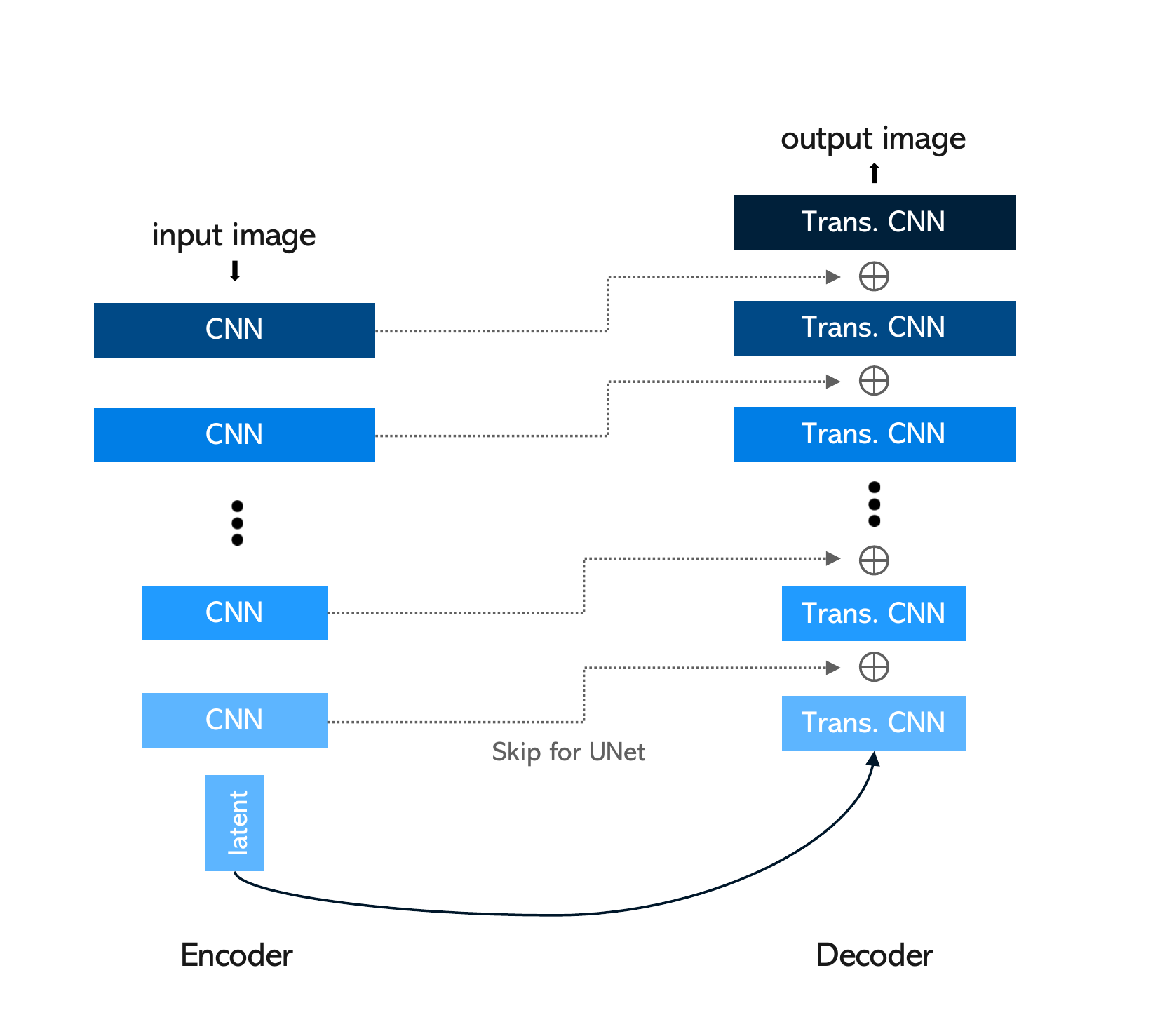}}
\caption{Architecture of symmetric auto-encoder (AE) and U-Net}
\label{f2}
\end{figure}
Focusing on the most successful architecture proposed for image processing in \cite{UNet:15}, we exploit the UNet architecture for our prediction models to improve prediction accuracy. 
Figure \ref{f2} depicts the UNet architecture that is a convolutional AE with skip connections. 
A convolutional AE consists of a convolutional encoder on the left-hand side of the figure and a transposed convolutional (a.k.a up-convolution) decoder on the right-hand side.
The encoder compresses the input image into a latent vector (code in a latent space), and the decoder constructs the output image from the code generated by the encoder. 
Colored blocks in the same color represent two convolutional neural networks (CNNs) with the same feature size that can be concatenated with each other by a skip connection.
As shown in the figure, this UNet uses a symmetric AE architecture with skip connections between each layer in the encoder and decoder, which will be referred to as {\it full UNet} in this paper.
In comparison, UNet based on an asymmetric AE will be called {\it partial UNet} hereafter since an asymmetric AE allows skip connections only between the mirrored layers in the encoder and decoder stacks. 
In the simulation section, we will show that the introduction of skip connections to the standard AE by UNet has a profound impact on prediction performance. 

\begin{figure}
\centerline{\includegraphics[trim=0cm 0.5cm 1cm 0.0cm,clip,width=0.80 \linewidth]{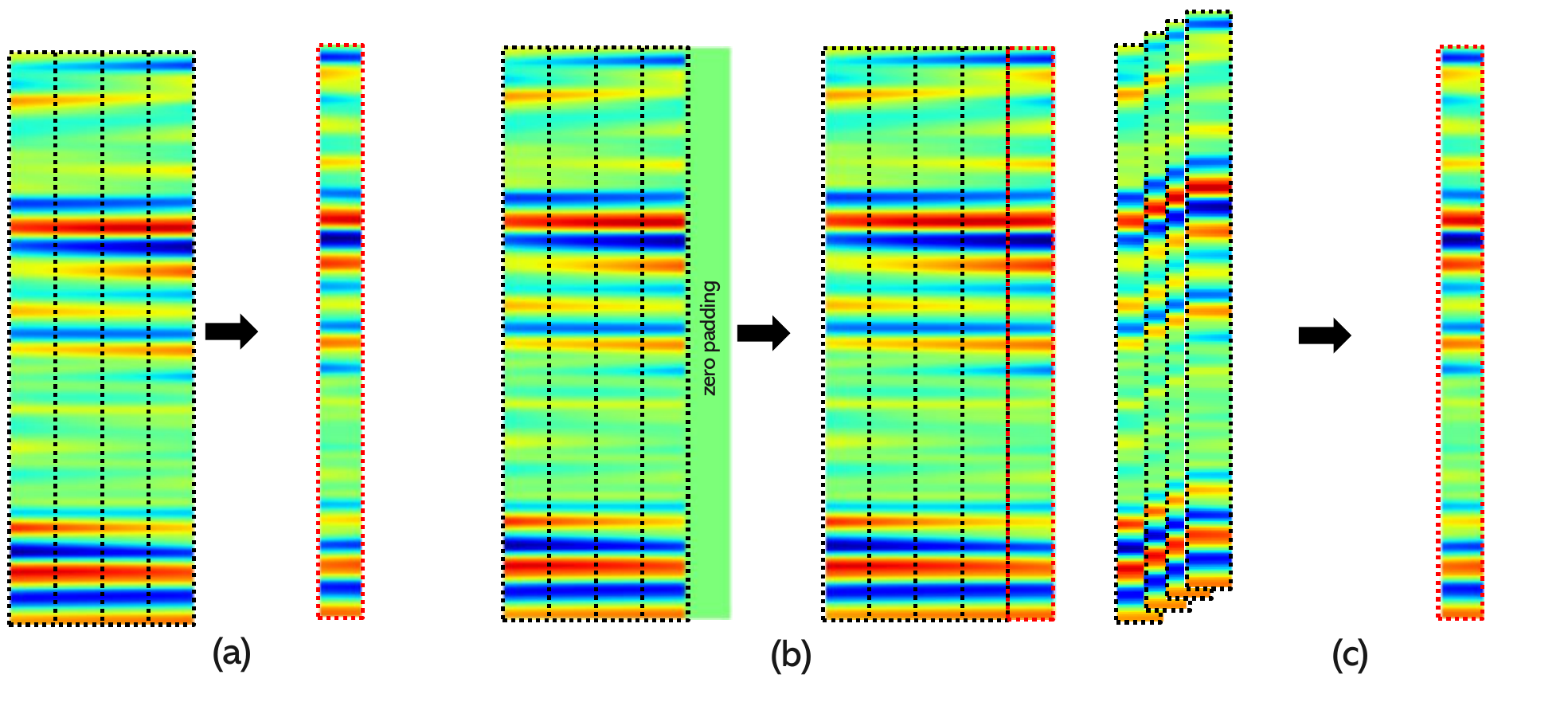}}
\caption{Conditioning input and target output image pairs: (a) baseline model, (b) image completion model, and (c) next-frame prediction model}
\label{f3}
\end{figure}
All the three models can be trained by using the same datasets but each model uses different input and output image pairs based on the same sample $s_n=\left[h^n_{k-m+1} ,\cdots, h^n_{k},h^n_{k+1}\right]$. Figure \ref{f3} illustrates different input and output image pairs for three models $\bff_{\theta}^{bs}$, $\bff_{\theta}^{ic}$, and $\bff_{\theta}^{nf}$, where the conditioning channel images, composed from the past four slots $k-3$ to $k$, are depicted by a black box and the target channel image, given by next slot $k+1$, is depicted by a red box.
As a naive baseline, Figure \ref{f3}-a) illustrates the input and output image pairs for the baseline model $\bff_{\theta}^{bs}$.
The baseline model $\bff_{\theta}^{bs}$ uses an asymmetric AE or a partial UNet to predict a small output image from a large input image that minimizes the loss function in \eqref{eq:l1}. 
The channel images from the four conditioning state images $h_{k},h_{k-1},...,h_{k-3}$ is combined into a single image and used as one large conditioning input image for $\bff_{\theta}^{bs}$.
A small output image is readily defined by a ground true channel image from the target state $h_{k+1}$.
$\bff_{\theta}^{bs}$ uses this asymmetric AE architecture to perform the required down-sampling by the encoder and up-sampling operation by the decoder to transform the large image input into the small image output. 

In order to use a symmetric AE or full UNet, the input and output images need to have the same size. 
As shown in Figure \ref{f3}-b), we can define the input and output pairs for the image completion model $\bff_{\theta}^{ic}$ by applying a zero padding to the input image.
All the five images of $h_{k-3}$ to $h_{k+1}$ are combined into a single image and used as a target output image for $\bff_{\theta}^{ic}$.
To match the spatial size between the input and the output image, the same input image as in $\bff_{\theta}^{bs}$ is concatenated with a zero padding size $T\times F$ and used as a new input image for $\bff_{\theta}^{ic}$.
The part of zero padding in the corresponding input image is depicted by a part of all zeros (green) in the right hand side of Figure \ref{f3}-b). 

We note that the baseline and image channel completion models are implemented with the first layer of a single {\it CNN channel}.
In order to make maximum use of the GPU memory, we can take the four sequential input images into four parallel {\it CNN channels}, which leads to next-frame prediction $\bff_{\theta}^{nf}$ (a.k.a. forward video prediction). As shown in Figure \ref{f3}-c), 
accommodating the past channel images via parallel CNN channels simultaneously, the prediction model $\bff_{\theta}^{nf}$ is trained to predict future channel image of the target state $h_{k+1}$ while all the input and output images have the same size $T\times F$.

\section{Numerical results} \label{sec:3}
In this section, we present empirical evaluations to demonstrate the performance benefit of data-driven approach to channel prediction on both synthetic and real channel data. 

We can also solve for the function $\bff$ in (\ref{eq:pre}) using Kalman filter.
We developed the KF-based predictor using a linear stochastic model called autoregressive (AR) with the order $p$ which serves as a linear model for (\ref{eq:pre}). 
The underlying assumption is that the channel in the future can be modelled by a linear combination of $p$ previous values with a process noise and predicted by Kalman filter.
In our numerical experiment, AR model parameters are estimated by using the Yule-Walker equation based on channel data from the previous 15 SRS transmissions and the estimated AR model is used to predict channel in a next SRS transmission. 
For more details on the Kalman filter design, we suggest the reader to refer to \cite{RikkeThesis:18}. 

\begin{table}
\caption{Comparison between prediction models}
\begin{center}
\begin{tabular}{|c|c|c|c|}
\hline
\textbf{models} & \textbf{\textit{architecutre}}& \textbf{\textit{input and output image sizes}}& \textbf{\textit{$l_1$}} \\
\hline
image & symetric AE & $\!(m\!+\!1\!)T\!\times \! F \! \rightarrow \! (m \! + \! 1)T \! \times \! F \!$ & $ \! \! 0.1263 \! \! $\\
\cline{2-4} 
completion & full UNet & $\!(m\!+\!1\!)T\!\times \! F \! \rightarrow \! (m \! + \! 1)T \! \times \! F \!$ &$ \! \! 0.0057 \! \! $\\
\hline
baseline & asymetric AE & $mT \!\times \! F \rightarrow T \!\times \! F$ &$ \! \! 0.1189 \! \! $\\
\cline{2-4}
model & partial UNet & $mT \!\times \! F \rightarrow T \!\times \! F$ &$ \! \! 0.0032 \! \! $\\
\hline
next-frame & symetric AE & $T\!\times \! F \rightarrow T \!\times \! F$ & $ \! \! 0.1172 \! \! $\\
\cline{2-4} 
prediction & full UNet & $T \!\times \! F \rightarrow T \!\times \! F$ & $ \! \! 0.0023 \! \! $\\
\hline
\end{tabular}
\label{t1}
\end{center}
\end{table}

We implemented the three models $\bff_{\theta}^{bs}$, $\bff_{\theta}^{ic}$, and $\bff_{\theta}^{nf}$ in TensorFlow 2 and trained them using a synthetic dataset with 5,000 samples and a real dataset with 150,000 samples. 
The difference in data size with the synthetic data having far fewer data samples than the real dataset comes from the fact that we can generate synthetic data with varied channel characteristics by randomly choosing the channel parameters at the beginning of each drop in simulation unlike the real world system in which these channel parameters tend to stay relatively constant through measurement. 
As shown in Figure \ref{f2}, AE and UNet share a common encoder and decoder structure except skip connections are only used for UNet. 
Each of encoder and decoder consists of eight convolution layers for all the three models but additional two input layers are applied for the baseline model in order to accommodate different input and output sizes.
For fair comparison, the three models are designed with approximately the same number of parameters ($\approx$ one million parameters). 
The convolutional layers in the encoder use LeakyReLU as an activation function. 
On the other hand, the transposed convolutional layers in the decoder use ReLU except the last layer uses tanh.
All the layers in the encoder and decoder are followed by batch normalization prior to the activation function except the input and output layer.
Dropout layers are added for the first three layers of the decoder in order to prevent overfitting of the model. 

First, we evaluate the UNet-based three models on synthetic channel data in comparison with the AE-based models in mean absolute errors.
The prediction results on the synthetic dataset are summarized in Table \ref{t1}, where the three models are listed in order of decreasing input and output image size.
We can see from the results that the UNet architecture yields significantly lower prediction error than the AE architecture for each model, demonstrating superior accuracy over AE. 
The results in the table show that for the next frame prediction model which has the smallest complexity UNet can improve channel prediction performance up to 50 times in mean absolute errors, compared to the AE.

\begin{figure}
\centerline{\includegraphics[trim=0.7cm 1.2cm 0.2cm 0.55cm,clip,width=0.87 \linewidth]{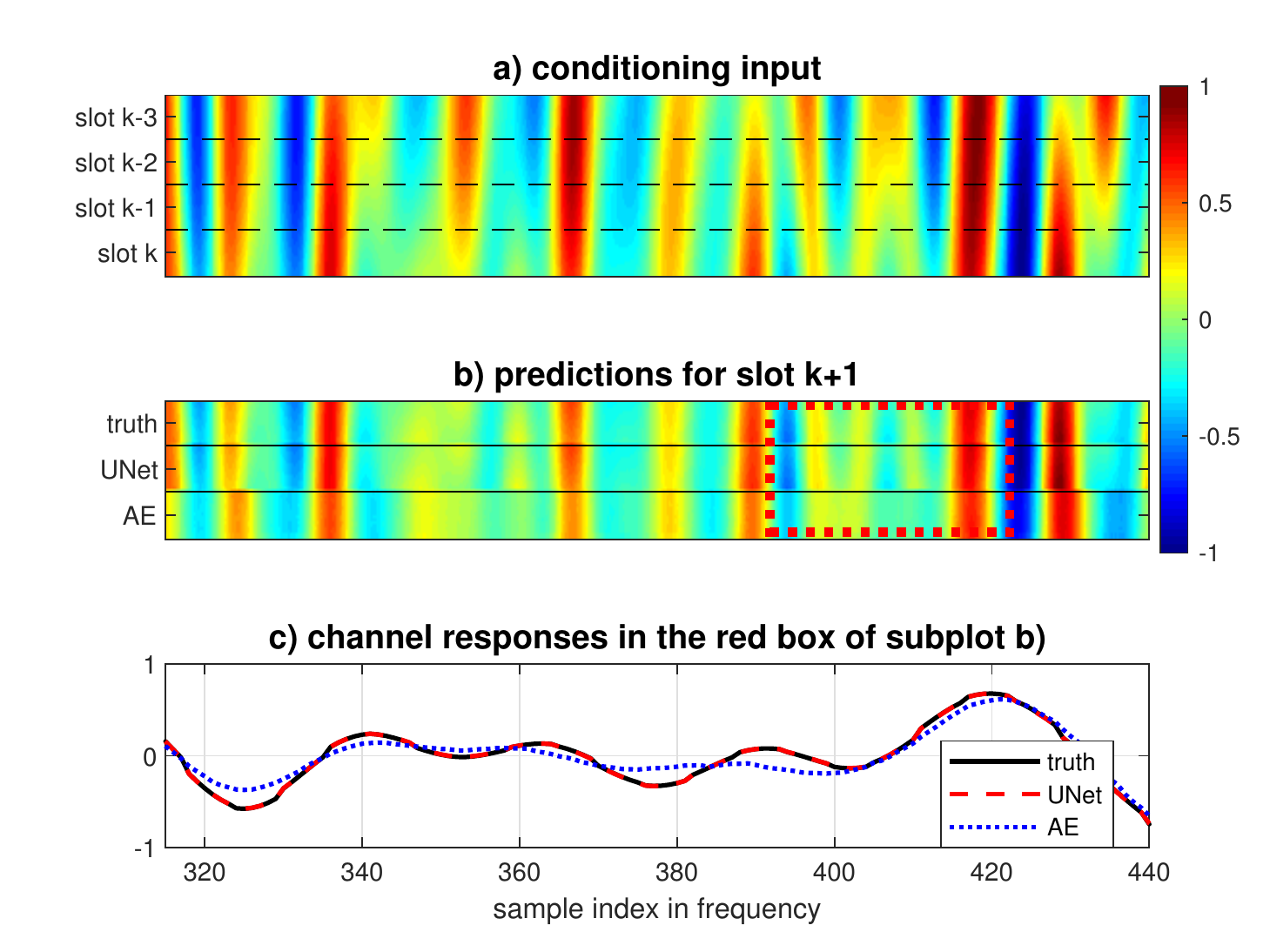}}
\caption{An illustrative example of channel prediction}
\label{f4}
\end{figure}

\begin{figure}
\centerline{\includegraphics[trim=0.7cm 1.2cm 0.2cm 0.55cm,clip,width=0.87\linewidth]{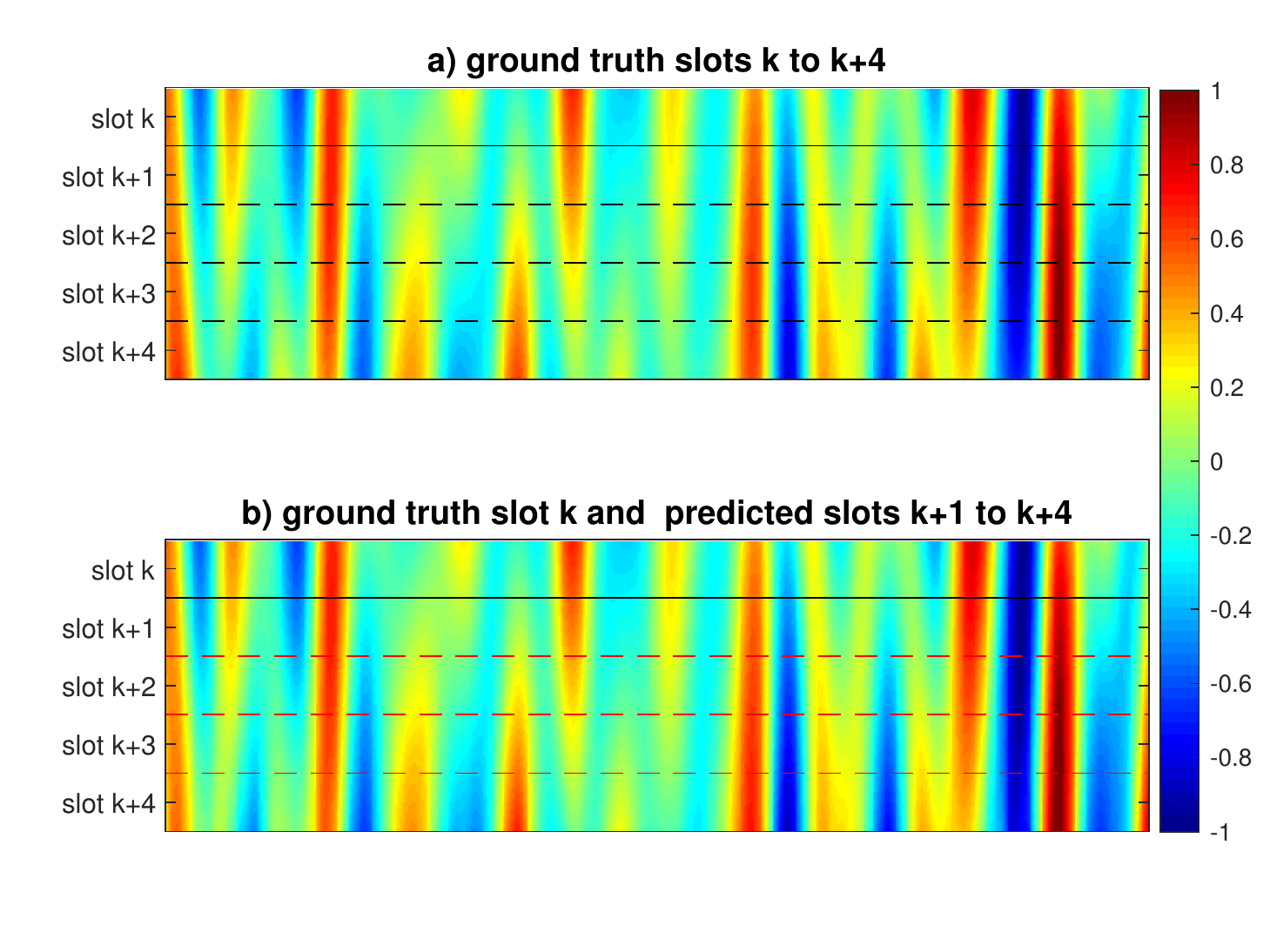}}
\caption{An illustrative example of four-step ahead channel prediction}
\label{f5}
\end{figure}
Figure \ref{f4} illustrates channel prediction by the UNet-based next frame prediction, compared to the AE-based next frame prediction and the ground truth.
The channel images and corresponding numerical results confirm superior performance of the UNet over the AE.
The numerical results for the three models designed with five million parameters, while not presented here, show the same general trend in performance as found in Table \ref{t1}.

Equation \eqref{eq:pre2} shows that our prediction models can be applied for multi-step ahead channel predictions on a slot basis in autoregressive manner. 
At each step $\bar{k}$, the new channel state $h_{\bar{k}+1}$ is predicted based on the last $m=4$ channel states $h_{\bar{k}},h_{\bar{k}-1},...,h_{\bar{k}-3}$. 
Figure \ref{f5} shows the four-step ahead prediction by the UNet-based next-frame prediction model for the steps $\bar{k}=k,k+1,k+2,k+3$ to generate the channel predictions at slots $k+1,k+2,k+3,k+4$ autoregressively.
The results demonstrate that the UNet-based next-frame prediction model can successfully predict channels up to four steps ahead given past four slot channels 

For validation on real channel data, in the rest of this section we will evaluate the UNet-based next-frame prediction by using real channel data obtained from the measurement campaign. 
We assume to use past four SRSs $k-3$ to $k$ as a conditioning input image for real data to predict next SRS $k+1$ . 
We compare three different CSI assumptions including perfect CSI that is defined by true channel at next SRS $k+1$, aged CSI that is given by true channel at last SRS $k$, and predicted CSI that is given by predicted channel at next SRS $k+1$ based on the past four SRSs.
In order to verify the accuracy, we also compare our DL approach with the KF approach. 
For a fair comparison with the proposed prediction models, the AR-order $p$ is fixed to be equal to the memory size $m$.

\begin{figure}
\centerline{\includegraphics[trim=0.5cm 0.2cm 0.5cm 0.5cm,clip,width=0.807\linewidth]{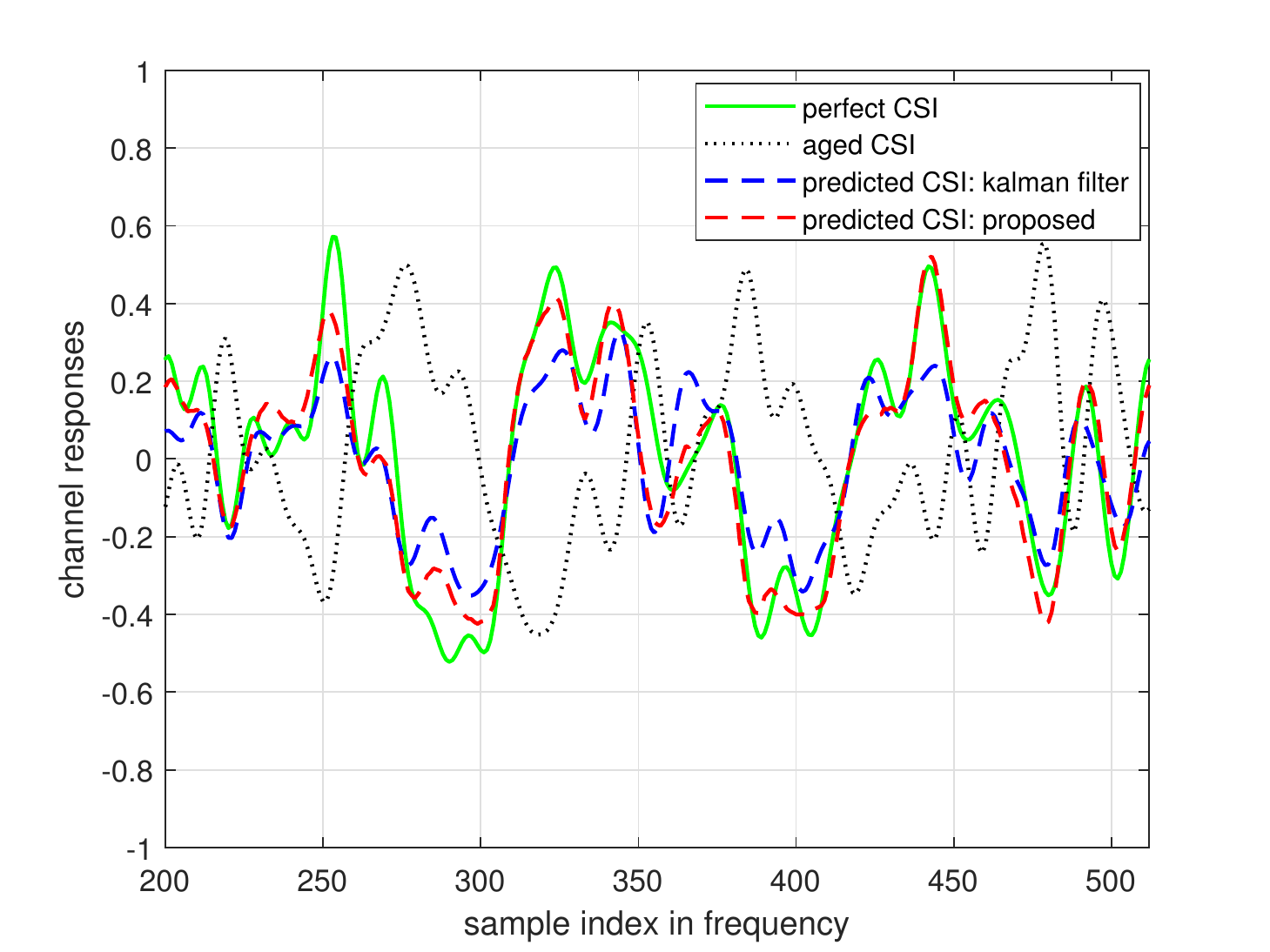}}
\caption{A representative example of channel prediction}
\label{f6}
\end{figure}
\begin{figure}
\centerline{\includegraphics[trim=0.5cm 0.2cm 0.5cm 0.5cm,clip,width=0.807\linewidth]{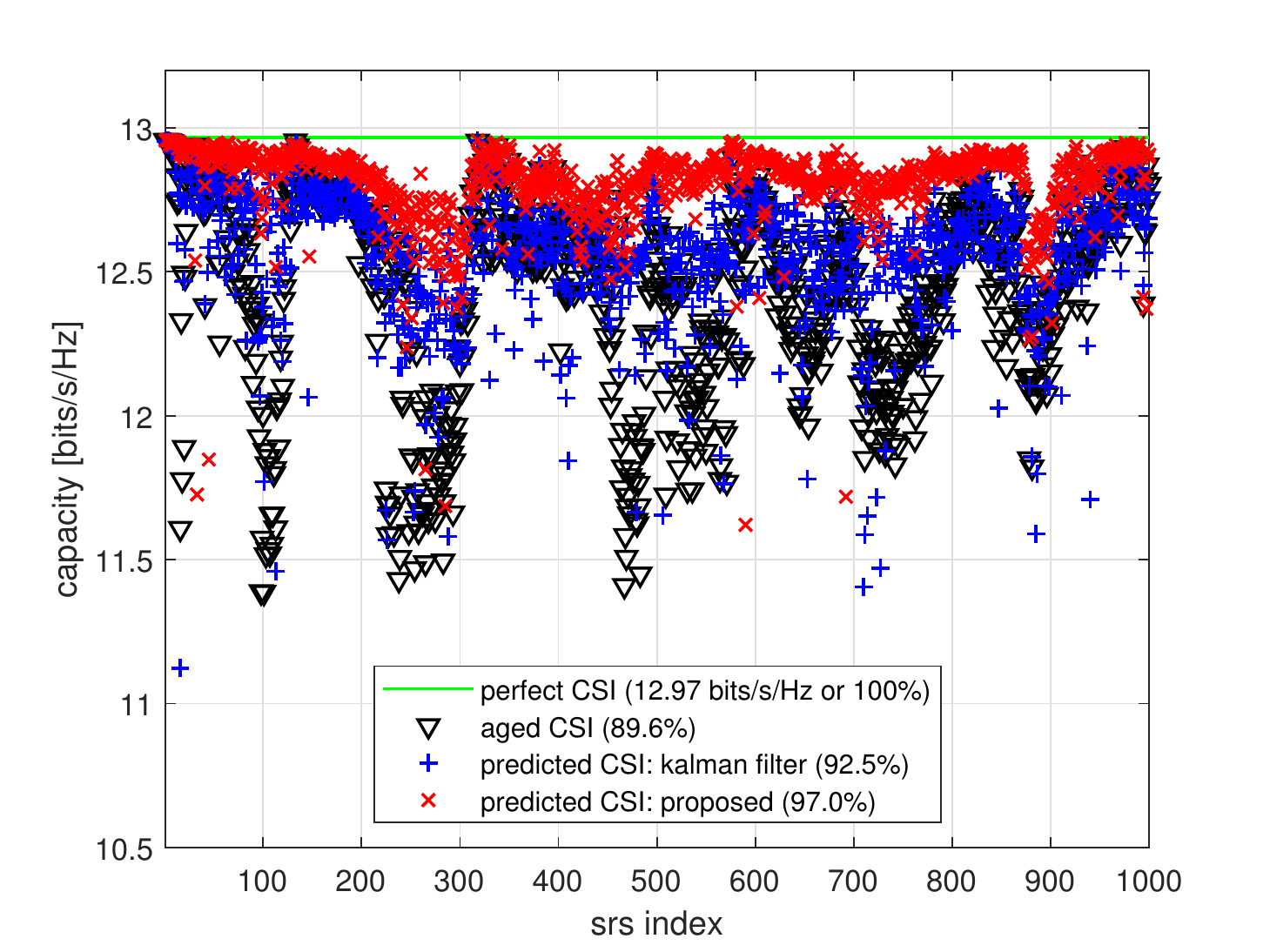}}
\caption{10\% outage capacity over the measurement trajectory}
\label{f7}
\end{figure}
Our experiment results show that the next-frame prediction model gives the mean absolute error $l_1=0.0705$, achieving 52\% lower prediction error than the mean absolute error $l_1=0.1466$ by the KF approach.
Figure \ref{f6} shows a representative example of channel prediction by the proposed approach, compared to the KF approach. This particular example allows for both the proposed and traditional approach to achieve an instantaneous  prediction accuracy close to its average. 
The simulation results demonstrate that the proposed DL approach can predict channel responses that agree fairly well with the ground true channel responses whereas the predicted CSI by the KF approach exhibits a significant gap to the ground true channel response, especially relatively large mismatches around the peaks of channel response curves.

The performance comparison of the prediction models can be made in terms of capacity as channel capacity degrades with aged CSI. 
In this work, we use channel capacity and evaluate its outage capacity of a rank-1 precoding scheme over 8-by-1 MISO spatial channel. 
The $\epsilon$-outage capacity is defined as maximum rate below which reliable transmission is possible at outage probability of $\epsilon$ ($p_{out}$) 
\bea
C_{\epsilon}=\max_{r}\left\{p_{out}\left[ \log_{2}\left(1+\rho\left|\vc_{(t,f)}\vw_{(t,f)}\right|^2 \right)\leq r \right] \leq \epsilon \right\}\!, \!\!\!
\eea
where $\rho$ is the SNR parameter, $\vc_{(t,f)} \in \dC^{1\times n_t}$ denotes the complex-valued channel vectors at the $(t,f)$th element of ground truth, and $\vw_{(t,f)} \in \dC^{n_t\times 1}$ represents the precoding vector obtained by given CSI at the $(t,f)$th element. 

The maximum ratio transmission (MRT) precoding is optimal for this MISO setup by maximizing the signal gain and CSI at the transmitter is required to enable the precoding. 
We will compare the three different CSI assumptions including perfect CSI, aged CSI and predicted CSI, denoted by $\hat{\vc}^{tr}_{(t,f)}$, $\hat{\vc}^{ag}_{(t,f)}$, and 
$\hat{\vc}^{pr}_{(t,f)}$, respectively.
The MRT solution is given by $\vw_{(t,f)}^{mrt} =\hat{\vc}_{(t,f)}^{\dag}$, where $\left(\cdot\right)^{\dag}$ denotes the conjugate transpose of vector and $\hat{\vc}_{(t,f)} \in \left\{\hat{\vc}^{tr}_{(t,f)}, \hat{\vc}^{ag}_{(t,f)},\hat{\vc}^{pr}_{(t,f)} \right\}$.

In Figure \ref{f7}, we have evaluated 10\% outage capacity performance ($\epsilon$=0.1) under the three different CSI assumptions in order to quantify impact of channel prediction accuracy on precoding performance.
The results show that the MRT precoding with the aged CSI suffers about 10.3\% performance loss due to channel aging with respect to the ideal case given by perfect CSI.
As shown in Figure \ref{f7}, the performance degradation caused by channel aging can be reduced by applying predicted CSIs.
Figure \ref{f7} shows that the UNet-based prediction model $\bff_{\theta}^{nf}$ can reduce the performance loss by 71\%, which corresponds to 97\% of the maximum capacity with perfect CSI, while the traditional KF approach only shows a 27\% reduction.

\section{Conclusion} \label{sec:4}
In this paper, we have proposed and experimentally evaluated a data-driven deep learning approach to channel prediction problem. 
The data-driven approach has been proven to be successful in predicting accurately temporal evolution of radio channels without knowledge of the underlying channel dynamics.
Meanwhile, AE and UNet are compared against each other for each prediction model in order to light the importance of UNet model in deep learning-based channel prediction.
We have also evaluated the proposed data-driven approach in terms of outage capacity in order to quantify impact of channel prediction accuracy on precoding performance.
The proposed data-driven channel prediction models and results in this paper open a new opportunity toward intelligent radio access network, especially in radio resource  and beam management. 

\section*{Acknowledgment}
This work has been partly funded by the European Commission through the H2020 project Hexa-X (Grant Agreement no. 101015956).

\bibliographystyle{ieeetr}
\bibliography{REFbib}

\end{document}